\documentclass[aps,prl,showpacs,twocolumn,superscriptaddress]{revtex4}
\usepackage{bm,color,amsmath,amssymb,mathrsfs,latexsym,graphicx,psfrag}

\newcommand{\beq}{\begin{equation}}
\newcommand{\eneq}{\end{equation}}

\begin{document}

\title{Screening Behavior and Scaling Exponents from Quantum Hall Wavefunctions}

\author{B. Andrei Bernevig}
\affiliation{Department of Physics, Princeton University, Princeton, New Jersey 08544, USA}
\author{Parsa Bonderson}
\affiliation{Station Q, Microsoft Research, Santa Barbara, California 93106-6105, USA}
\author{Nicolas Regnault}
\affiliation{Department of Physics, Princeton University, Princeton, New Jersey 08544, USA}
\affiliation{Laboratoire Pierre Aigrain, ENS and CNRS, 24 rue Lhomond, F-75005 Paris, France}

\begin{abstract}
We provide a robust and generic method to assess the screening properties and extract the scaling exponents of quasiparticle edge excitations of quantum Hall states from model wavefunctions. We numerically implement this method for the fundamental quasihole and hole excitations of several model states. For the Laughlin, Moore-Read, and $\mathbb{Z}_3$-Read-Rezayi states, we find agreement with the predicted edge theory, verifying the bulk-edge correspondence. We also use this to obtain the first clear microscopic demonstration of the pathologies of the Gaffnian wavefunctions.
\end{abstract}

\date{\today}


\maketitle

Since Laughlin's explanation~\cite{laughlin83prl1395} of the fractional quantum Hall (FQH) effect~\cite{Tsui82} in terms of incompressible quantum liquids, model (trial) wavefunctions have played a significant role in understanding FQH states. This understanding has benefited greatly from the realization~\cite{fubini1991,Moore-91npb362} that conformal field theories (CFTs)~\cite{DSM} can generate model FQH wavefunctions (including Laughlin's) using conformal blocks of appropriately chosen CFT operators. The analytic properties of such wavefunctions are manifested by the CFTs from which they are produced, allowing straightforward extraction of important experimentally measurable quantities characteristic of the universality classes which these wavefunctions (are intended to) represent. Such quantities include the charges and braiding statistics of quasiparticles, as well as the edge modes' scaling exponents and chiral central charge, which are measurable, for example, through tunneling, interferometry, and thermal transport experiments~\cite{wen-92jmp1711,Bishara09,Read-00prb10267}.

However, the validity of a CFT based FQH state and its experimental predictions depends upon the assumption that it produces legitimate FQH wavefunctions whose inner products (with respect to particle coordinates) match those of the CFT. This assumption has been verified for the Laughlin~\cite{laughlin83prl1395} and Moore-Read (MR)~\cite{Moore-91npb362} states by mapping their respective wavefunctions' inner products to the partition functions of certain classical two-dimensional plasmas~\cite{laughlin83prl1395,Bonderson11b} that are in their metallic phases~\cite{Caillol82,Herland12}.
More generally, plasma mappings can be constructed for proposed FQH states based on all the minimal model CFTs~\cite{Bonderson11b} and their $\mathbb{Z}_{k}$ generalizations~\cite{BondersonWIP}, including the Read-Rezayi (RR) states~\cite{Read-99prb8084}, but the corresponding plasmas are more complicated and their screening properties not yet well-established. Hence, the question of which CFTs produce legitimate FQH wavefunctions remains largely unsettled.

One may also consider states generated using non-unitary CFTs, such as the ``Gaffnian'' (Gf) state~\cite{yoshioka88prb3636,simon-07prb075317} generated from the $\mathcal{M}(5,3)$ non-unitary minimal model CFT [together with a U$(1)$ Laughlin-like charge sector]. It is difficult to envision how FQH wavefunctions could emerge from non-unitary CFTs free of pathologies, such as non-unitary edge modes and braiding statistics or gaplessness in the bulk, and arguments have been made~\cite{simon-07prb075317,read09prb045308,read09prb245304} that they cannot describe gapped phases. However, the failures of such states can be subtle and a complete microscopic understanding of their deficiencies is still missing. Numerical studies of Gf wavefunctions for small system sizes produce some indications that the Gf could be an acceptable FQH state~\cite{simon-07prb075317,Toke-09prb205301,bernevig-08prl246806,regnault_PhysRevLett.101.066803}, while the first signs of problems appear in numerical studies of the entanglement spectra~\cite{regnault-09prl016801}, which suggest non-unitarity in the edge modes, and of states with quasiholes~\cite{Toke-09prb205301}, which exhibit indications of level crossings when deforming between the Coulomb and Gf Hamiltonians. However, these results give no clear indications of the behavior in the thermodynamic limit and clear demonstrations of the anticipated failures of the Gf wavefunctions have not been identified.

In this paper, we provide methods of analyzing model wavefunctions of CFT generated states that can help determine whether they exhibit the required properties of FQH states and verify the bulk-edge correspondence, which conjectures that the $(1+1)$D edge theory is described by the same CFT used to generate the $2$D bulk wavefunctions, from microscopic computations. We devise a numerical method of extracting the scaling exponents of edge excitations from microscopic wavefunctions, using Jack polynomial expansions to produce sequences that must converge in the thermodynamic limit to the exponents for a properly screening state. Applying these methods, we find behavior consistent with the respective CFT descriptions for the Laughlin, MR, and RR states (though system size limitations prevent a clean extraction of the RR state's quasihole exponent). We analyze only bosonic FQH states here, but the generalization to fermionic states is straightforward and yields similar results~\cite{BernevigWIP}. We find the behavior for the Gf wavefunctions to be inconsistent with proper screening and the CFT description of a FQH state, providing clear indications of their failure.

The basic configuration that we use to study candidate states is the planar disk geometry with some number of quasiparticles, one of which is located at (complex) coordinate $\eta$, while all others are located at the origin of the disk. The corresponding (unnormalized) wavefunction for $N$ bosonic particles with coordinates $z_1,\ldots, z_N$ can generally be written in the form
\begin{equation}
\label{qpwavefunction}
\Psi(\eta ;z_i)= \sum_{a=0}^{n_{\phi}} \eta^a P_{a} (z_1,\ldots,z_N) \,\, e^{-\frac{1}{4 \ell^2} \sum\limits_{i=1}^{N} \left|z_i \right|^2 }
,
\end{equation}
where $P_{a} (z_1,\ldots,z_N)$ are symmetric polynomials, $n_{\phi}$ is the highest power of $\eta$, and $\ell=\sqrt{\hbar c / eB}$ is the magnetic length. The inner product of the model wavefunction with its quasiparticle at possibly different positions is
\begin{equation}
\label{eq:G}
\Gamma(\bar{\eta}, \eta^{\prime}) = \int \prod_{j=1}^N  d^2 z_j \overline{\Psi(\eta ; z_i)} \Psi(\eta^{\prime} ; z_i)
= \sum\limits_{a=0}^{n_{\phi} } (\bar{\eta} \eta^{\prime})^a \mathcal{N}_a
\end{equation}
where we have defined
\begin{equation}
\mathcal{N}_a \equiv \int \prod_{j=1}^N  d^2 z_j  \left|  P_{a} (z_1,\ldots,z_N) \right|^2 \,\, e^{-\frac{1}{2 \ell^2} \sum\limits_{i=1}^{N} \left| z_i \right|^2 }
\end{equation}
to be the norm-squared of $P_{a} (z_1,\ldots,z_N)$, and used the fact that these polynomials with different $a$ are orthogonal, since their total orders are unequal.

While a quasiparticle cannot physically exist outside of the quantum Hall droplet, one can \emph{formally} write model wavefunctions with the quasiparticle coordinates located anywhere.
The configuration with a quasiparticle outside the Hall droplet can be used to assess the candidate state's screening properties, extract the quasiparticle's scaling exponent, and determine whether the bulk-edge correspondence holds, using arguments similar to those applied to Laughlin states by Wen~\cite{wen-92jmp1711}. If the state screens properly, the norm of the wavefunction with the quasiparticle outside the disk should take the form
\begin{equation}
\label{eq:norm}
\left\| \Psi \right\|^2 = \Gamma(\bar{\eta}, \eta) \sim C \left| \eta \right|^{2 n_{\phi} } \left(1-\frac{R^2}{\left| \eta \right|^2} \right)^{-g}
\end{equation}
in the thermodynamic limit ($N \rightarrow \infty$) for $\left| \eta \right| - R \gg \ell$, where $C$ is an overall constant, $R$ is the radius of the disk of Hall fluid, and the exponent $g$ is a constant that depends on the state and quasiparticle. This form can be justified for certain model states (including the ones studied here) by mapping the norm to the free energy of a classical 2D plasma~\cite{laughlin83prl1395,Bonderson11b,BernevigWIP}. In its screening phase, the plasma behave like a metal, so the free energy of the system is the Coulombic energy of a metallic disk with a point charge outside it. This can be computed by the method of images, which requires an image charge at $R^2 / \bar{\eta}$ that gives rise to the important $\left(1-\frac{R^2}{\left| \eta \right|^2} \right)^{-g}$ term in Eq.~(\ref{eq:norm}). Analytic continuation of Eq.~(\ref{eq:norm}) gives
\begin{eqnarray}
\label{eq:G_expansion}
\Gamma(\bar{\eta}, \eta^{\prime}) &\simeq & C \left( \bar{\eta} \eta^{\prime} \right)^{n_{\phi} } \sum\limits_{n=0}^{n_{\phi}} \binom{g+n-1}{n} \left(\frac{R^2}{\bar{\eta} \eta^{\prime}}\right)^n \\
\label{eq:G_analytic}
& \sim & C \left( \bar{\eta} \eta^{\prime} \right)^{n_{\phi} } \left(1-\frac{R^2}{\bar{\eta} \eta^{\prime}} \right)^{-g}
,
\end{eqnarray}
where the approximation holds in the regime where $\left| \eta \right|, \left| \eta^{\prime} \right| \gg R$, and the thermodynamic limit gives an infinite sum that converges for $\left| \eta \right|, \left| \eta^{\prime} \right| > R$.

If the candidate state is properly screening and gives rise to a well-defined edge theory described by a CFT, then the inner product of states with a quasihole on the boundary of the Hall droplet should match the (equal time) quasihole correlator in the edge CFT. Specifically, for $\eta= Re^{i \theta}$ and $\eta^{\prime}=Re^{i \theta^{\prime}}$ in the 2D planar wavefunction, and $w= e^{\frac{t}{R} + i \theta}$ and $w^{\prime}=e^{\frac{t}{R} + i \theta^{\prime}}$ in the $(1+1)$D edge theory, one should have (up to overall phases)
\begin{equation}
\label{eq:bulk_edge}
\Gamma(\bar{\eta} , \eta^{\prime})  \sim \left\langle \Phi^{\dagger} \left(w^{\prime} \right) \Phi \left(w\right)  \right\rangle = \left( w^{\prime} - w \right)^{-2h}
\end{equation}
where $h$ is the conformal weight of the quasihole operator $\Phi$ in the edge CFT. Thus, if these properties hold for the candidate state, then $g$ in Eq.~(\ref{eq:G_analytic}) should equal the scaling exponent $g=2h$ of the quasiparticle, in the thermodynamic limit.

By matching powers of $\bar{\eta} \eta^{\prime}$ in Eqs.~(\ref{eq:G}) and (\ref{eq:G_expansion}), we find
\begin{equation}
\mathcal{N}_{n_{\phi} -n} \simeq C \binom{g+n-1}{n} R^{2n}
,
\end{equation}
relating the symmetric polynomials' norms to $g$. Dividing by the $n=0$ expression (which is just $C=\mathcal{N}_{n_{\phi}}$), we obtain the sequence of equations
\begin{equation}
\binom{g^{(n)}+n-1}{n} =  \frac{ \mathcal{N}_{n_{\phi} -n} }{ R^{2n} \mathcal{N}_{n_{\phi}} }
\end{equation}
defining $g^{(n)}$, an approximation of $g$ coming from the $(n_{\phi}-n)$th order term of the polynomial expansion.
This provides a robust method to numerically compute $g$ and provides a consistency check on the properties of the candidate state, since each $g^{(n)}$ must independently converge to the same value in the thermodynamic limit. Dealing with finite system sizes, the most relevant expansion terms outside the disk are those with $n$ small and the accuracy of the approximations will decrease for larger $n$. [The terms in Eq.~(\ref{qpwavefunction}) with $n$ large must accurately describe the regime where the quasiparticle is inside the disk, which exhibits different behavior.] Hence, we focus on the $n=1$ and $2$ expressions
\begin{eqnarray}
\label{eq:g1}
g^{(1)} &=& \frac{ \mathcal{N}_{n_{\phi} -1} }{ R^{2} \mathcal{N}_{n_{\phi}} } , \\
\label{eq:g2}
g^{(2)} &=& \left[ 2 \frac{ \mathcal{N}_{n_{\phi} -2} }{ R^{4} \mathcal{N}_{n_{\phi}} } + \frac{1}{4} \right]^{\frac{1}{2}} -\frac{1}{2}
.
\end{eqnarray}

The model wavefunctions of the states we focus on can be simply expressed in terms of Jack polynomials~\cite{bernevig-08prl246802}, which makes them easier to generate numerically~\cite{bernevig-09prl206801}.
The $(k,m)$ series of Jack states at filling factor $\nu=k/m$ includes the Laughlin~$=(1,2)$, MR~$=(2,2)$, $\mathbb{Z}_k$-RR~$=(k,2)$, and Gf~$=(2,3)$ states. The Jack states are comprised of symmetric polynomials that vanish as the $m$th power of the separation of $k+1$ particles approaching each other.

We consider a $(k,m)$ Jack state with $N$ particles (where $N$ is a multiple of $k$), a flux $\frac{k-1}{k}$ quasiparticle at the origin, and a fundamental (flux $1/k$) quasihole at position $\eta$. (One can alternatively view the quasiparticle at the origin as a composite of $k-1$ fundamental quasiholes.) In this case~\cite{bernevig-08prl246806}, $n_{\phi} = N/k$ and
\begin{equation}
\label{qhwavefunction}
P_{a}(z_1,\ldots,z_N) = (-k)^{\frac{N}{k}-a} J^\alpha_{\mu_a} (z_1,\ldots,z_N)
\end{equation}
where $J^\alpha_{\mu_a}$ denotes the $N$ particle Jack polynomial with Jack parameter $\alpha=-(k+1)/(m-1)$ for the $(k,m)$ series and root configuration (in terms of occupation numbers)
\begin{equation}
\mu_a = \left[ \left( 1,k-1,0^{m-2} \right)^{a}, 0, \left( k, 0^{m-1} \right)^{\frac{N}{k} -a} \right],
\end{equation}
e.g. $\mu_0 = [0,k,0^{m-1},k,0^{m-1},k,\ldots, 0^{m-1}, k,0^{m-1}, k]$.
For this configuration, we use $R = \sqrt{ 2 \left( \nu^{-1} N + 1 \right) } \ell$ for the radius of the Hall droplet, as defined by the (neutralizing background) disk of uniform charge density $\rho = \nu e / 2 \pi \ell^2$ whose total charge is $(N +\nu)e$, equal to that of all quasiparticles minus all particles. (Note: There is some ambiguity in how to define the radius of the Hall droplet, as it cannot be a sharply defined quantity. Alternative choices produce similar results, as their differences vanish in the thermodynamic limit.)

\begin{figure}
  \includegraphics[width=4.2cm]{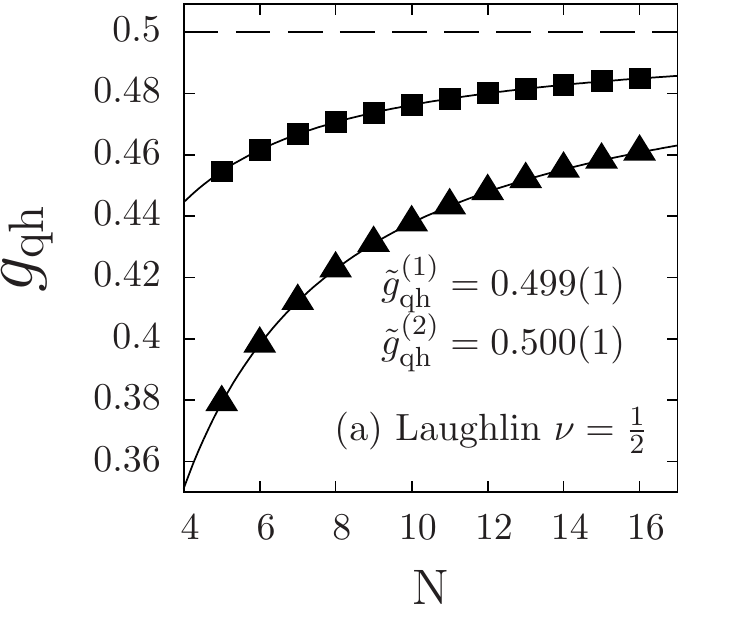}
  \includegraphics[width=4.2cm]{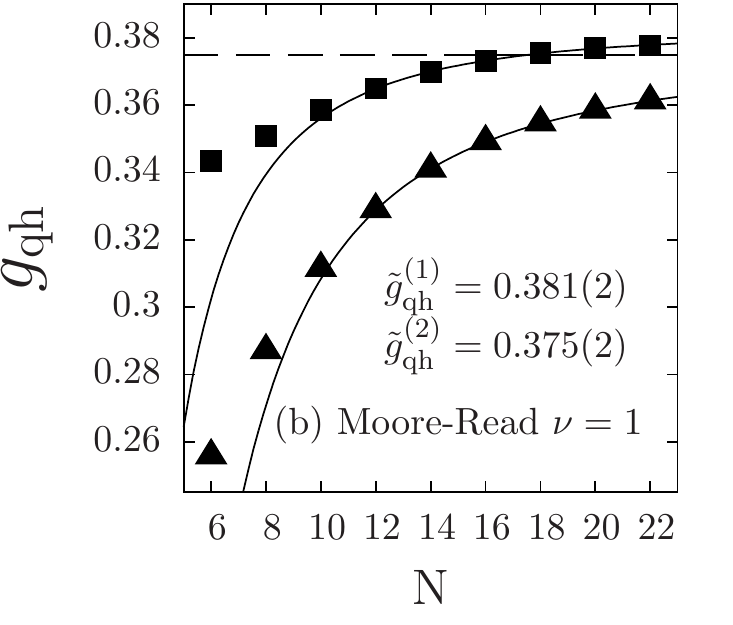}\\
  \includegraphics[width=4.2cm]{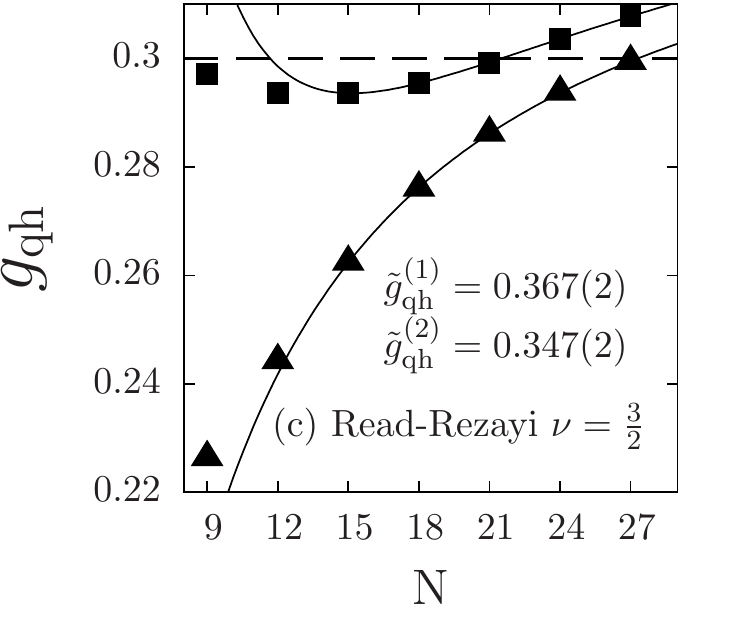}
  \includegraphics[width=4.2cm]{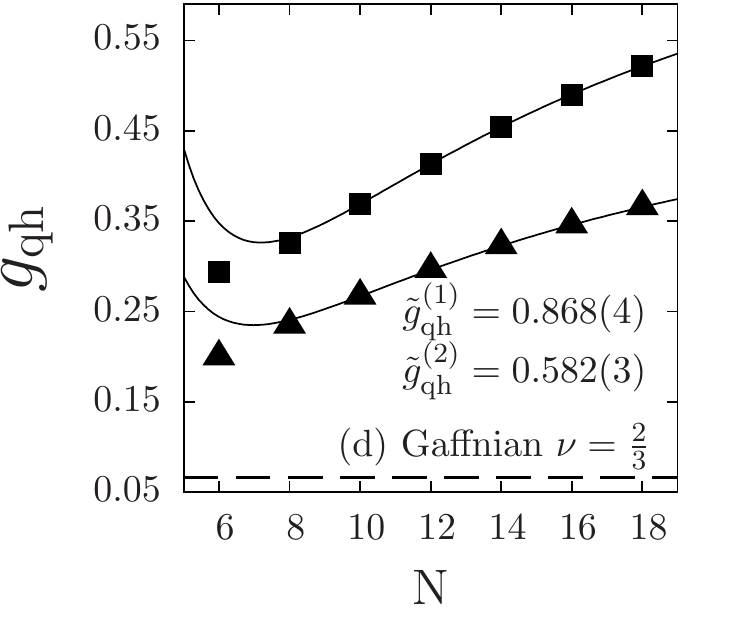}\\
  \caption{The fundamental quasihole scaling exponent versus the number of particles for the model states: (a) $\nu=1/2$ Laughlin, (b) $\nu=1$ MR, (c) $\nu=3/2$ $\mathbb{Z}_3$-RR state, and (d) $\nu=2/3$ Gf. $g_{\text{qh}}^{(1)}$ and $g_{\text{qh}}^{(2)}$ values are denoted using squares and triangles, respectively. The dashed lines indicate the values $g_{\text{qh}} = \frac{1}{2}$, $\frac{3}{8}$, $\frac{3}{10}$, and $\frac{1}{15}$ predicted from CFT for the respective states. The solid lines are quadratic fits in $1/N$, which give the $N\rightarrow \infty$ extrapolated values, $\tilde{g}_{\text{qh}}^{(n)}$.}
 \label{exponents_bosons_qh}
\end{figure}

The $\nu=1/m$ Laughlin state has the property $ \mathcal{N}_{n_{\phi} -1} = 2N \ell^2 \mathcal{N}_{n_{\phi}} $, which can be proven using $J^\alpha_{\mu_{n_{\phi} -1 }} = \sum_{i=1}^N z_i J^\alpha_{\mu_{n_{\phi} }}$, integration by parts, and $\sum_{i=1}^{N} \frac{\partial}{\partial z_i} J^{\alpha }_{\mu_{n_{\phi}}}=0$. This provides an exact expression $g_{\text{qh}}^{(1)} = (m + 1/N)^{-1}$, which rapidly converges to the expected value $g_{\text{qh}}=1/m$ for Laughlin quasiholes.

We numerically compute $g_{\text{qh}}^{(1)}$ and $g_{\text{qh}}^{(2)}$ for the Laughlin (up to $N=16$), MR (up to $N=22$), $\mathbb{Z}_3$-RR (up to $N=27$), and Gf (up to $N=18$) states, as shown in Fig.~\ref{exponents_bosons_qh}. These computations involve squeezed Hilbert spaces as large as $5.3 \times 10^9$. Motivated by the exact result for the Laughlin state, we use quadratic fits in $1/N$ to the $g^{(n)}$ points with $N \geq 5k$ in order to extrapolate the data to the $N\rightarrow \infty$ limit. The extrapolated values, $\tilde{g}_{\text{qh}}^{(n)}$, are indicated (with fitting errors) in the figure. For the Laughlin and MR states, $g_{\text{qh}}^{(1)}$ and $g_{\text{qh}}^{(2)}$ converge towards each other and produce excellent agreement with the predicted exponent values (within $2\%$). As $k$ increases, the behavior becomes worse. For the RR state, $g_{\text{qh}}^{(1)}$ and $g_{\text{qh}}^{(2)}$ converge towards each other, but produce $\tilde{g}_{\text{qh}}^{(n)}$ that differ from the predicted value by around $20\%$. We conjecture that this discrepancy is due to significant finite-size effects arising from the large size of the quasiparticle at the origin and that it will vanish as system size increases. In contrast to these states, $g_{\text{qh}}^{(1)}$ and $g_{\text{qh}}^{(2)}$ for the Gf state are very far from the predicted value and appear to be diverging from it, as well as from each other.

We note that Hu~\emph{et al.}~\cite{hu-10cm4716} proposed a different method to extract scaling exponents from quasiparticle tunneling between edges of an annulus. They found excellent agreement with the predicted values for the Laughlin-type (flux $1$) quasiholes of the Laughlin, MR, and RR states. However, their extracted exponents for the (non-Abelian) fundamental quasiholes of the MR and RR states exhibit a (possibly systematic, but so far unexplained) large disagreement with the predicted values.

\begin{figure}
  \includegraphics[width=4.2cm]{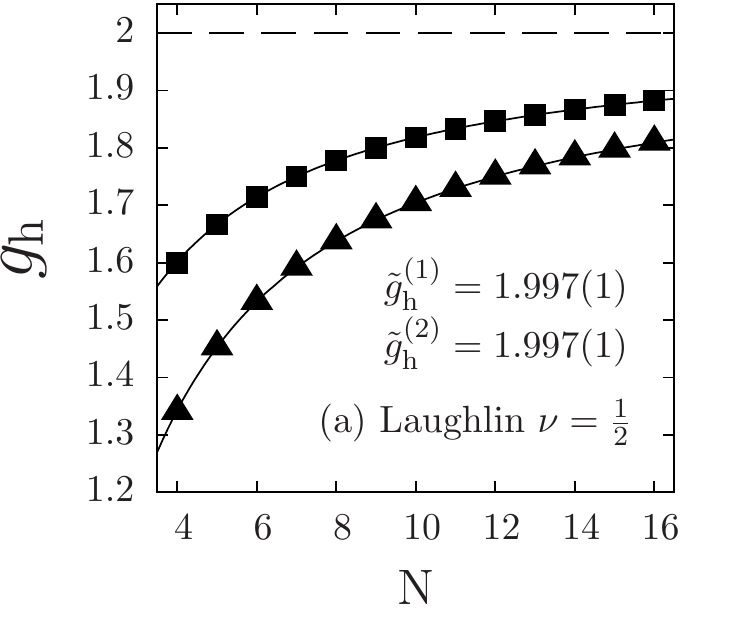}
  \includegraphics[width=4.2cm]{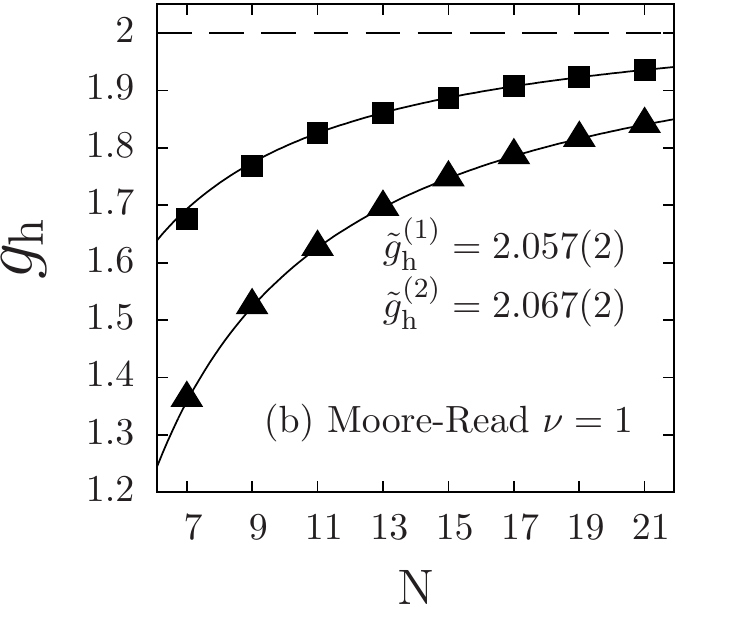}\\
  \includegraphics[width=4.2cm]{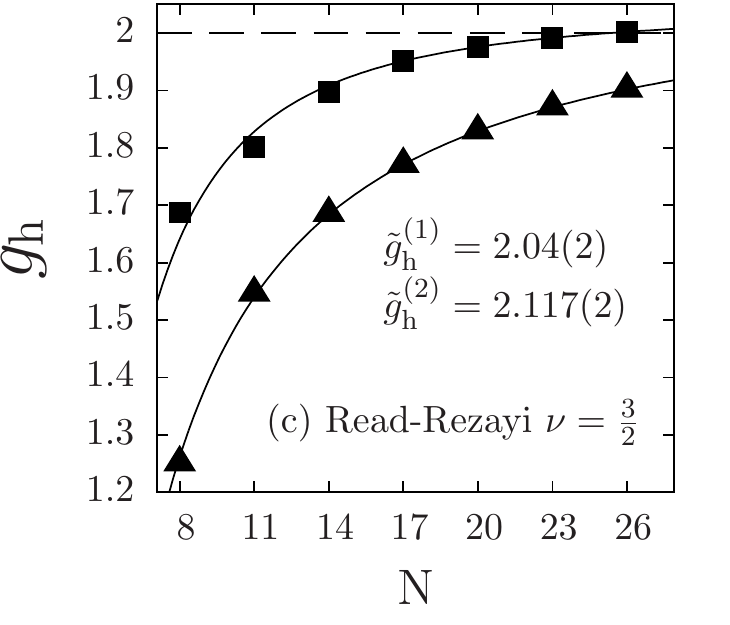}
  \includegraphics[width=4.2cm]{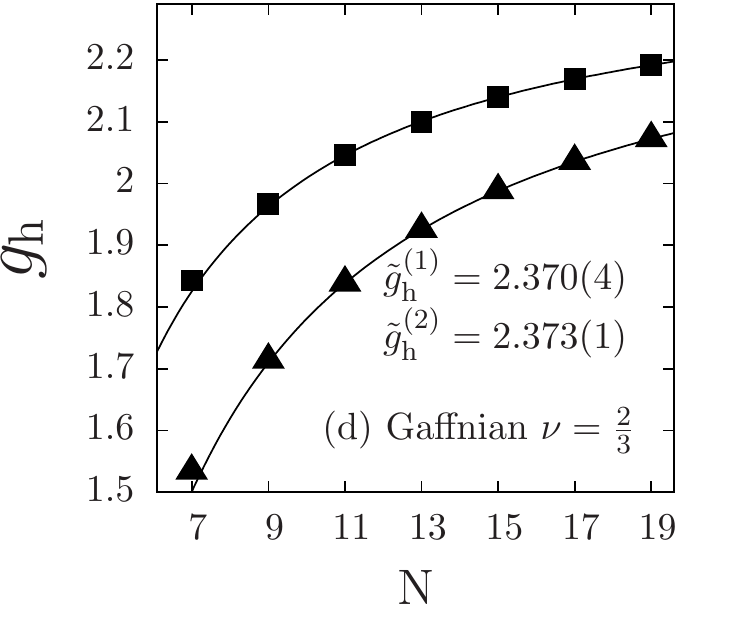}\\
  \caption{The hole (and particle) scaling exponent versus the number of (unpinned) particles for the model states: (a) $\nu=1/2$ Laughlin, (b) $\nu=1$ MR, (c) $\nu=3/2$ $\mathbb{Z}_3$-RR state, and (d) $\nu=2/3$ Gf. $g_{\text{h}}^{(1)}$ and $g_{\text{h}}^{(2)}$ values are denoted using squares and triangles, respectively. The dashed lines indicate the values $g_{\text{h}}=2$ predicted from the CFT for the unitary states. (CFT predicts $g_{\text{h}}=3$ for Gf.) The solid lines are quadratic fits in $1/N$, which give the $N\rightarrow \infty$ extrapolated values, $\tilde{g}_{\text{h}}^{(n)}$.}
 \label{exponents_bosons_el}
\end{figure}

We next consider a $(k,m)$ Jack state with $N$ particles (where $N+1$ is a multiple of $k$), with no quasiparticles at the origin, and a hole (i.e. a flux $\nu^{-1}$ quasiparticle with the same statistics as the particles) at position $\eta$. (One can alternatively view this state as the $N+1$ particle ground state $\Psi_0$ with one of its particles pinned at position $\eta$, i.e. $\Psi(\eta;z_1,\ldots,z_N) = \Psi_{0}(\eta,z_1,\ldots,z_N) e^{| \eta |^2 / 4 \ell^2}$.) In this case, $n_{\phi} =  \frac{m}{k} (N+1) - m$, $R = \sqrt{ 2 \nu^{-1} \left( N + 1 \right) } \ell$, and $P_{a}$ is generally equal to a sum of Jack polynomials, rather than a single Jack.

To obtain the polynomials $P_{n_\phi}, P_{n_\phi-1}, P_{n_\phi-2}$ needed to compute $g_{\text{h}}^{(1)}$ and $g_{\text{h}}^{(2)}$ for large sizes we require more refined Jack machinery and simply quote the results. In all cases, we have
\begin{eqnarray}
P_{n_\phi} &=& J^\alpha_{\lambda_{n_\phi}}, \qquad P_{n_{\phi}-1} = - \frac{m}{k} J^{\alpha}_{\lambda_{n_{\phi}-1}}, \notag \\
P_{n_\phi-2} &=& a_1  J^\alpha_{\lambda_{n_\phi-2}^{(1)}} + a_2  J^\alpha_{\lambda_{n_\phi-2}^{(2)}} , \notag \\
\lambda_{n_\phi} &=& \left[ (k,0^{m-1})^{\frac{N}{k}-1}, k-1 \right], \\
\lambda_{n_{\phi}-1} &=& \left[ (k,0^{m-1})^{\frac{N}{k}-2},k-1,1,0^{m-2}, k-1 \right], \notag \\
\lambda_{n_\phi-2}^{(1)} &=& \left[ (k,0^{m-1})^{\frac{N}{k}-2},k-1,0,1,0^{m-3}, k-1 \right], \notag \\
\lambda_{n_\phi-2}^{(2)} &=& \left[ (k,0^{m-1})^{\frac{N}{k}-3},(k-1,1,0^{m-2})^{2}, k-1 \right]
.
\notag
\end{eqnarray}
For the Laughlin, MR, $\mathbb{Z}_3$-RR, and Gf states, the coefficients in $P_{N_\phi-2}$ are given by $a_1=1$, $1$, $1$, and $-3/2$, respectively, and $a_2=14/5$, $5/7$, $26/81$, and $21/16$, respectively.

Similar to the quasihole case, the $\nu=1/m$ Laughlin state has the property $ \mathcal{N}_{n_{\phi} -1} = 2N \ell^2 \mathcal{N}_{n_{\phi}} $, which provides an exact expression $g_{\text{h}}^{(1)} = m (1+ 1/N)^{-1}$, which rapidly converges to the expected value $g_{\text{h}}=m$ for Laughlin holes/particles.

We numerically compute $g_{\text{h}}^{(1)}$ and $g_{\text{h}}^{(2)}$ for the Laughlin (up to $N=16$), MR (up to $N=21$), $\mathbb{Z}_3$-RR (up to $N=26$), and Gf (up to $N=19$) states, as shown in Fig.~\ref{exponents_bosons_el}. These computations involve squeezed Hilbert spaces as large as $1.5 \times 10^{10}$. We use quadratic fits in $1/N$ to the $g_{\text{h}}^{(n)}$ points with $N \geq 5k -1$ in order to extrapolate the data to the $N\rightarrow \infty$ limit. The extrapolated values, $\tilde{g}_{\text{h}}^{(n)}$, are indicated in the figure. For the Laughlin, MR, and RR states, we see that $g_{\text{h}}^{(1)}$ and $g_{\text{h}}^{(2)}$ converge toward each other and toward the value $g_{\text{h}}=2$ predicted from CFT, producing excellent agreement between $\tilde{g}_{\text{h}}^{(n)}$ and the predicted value. For the RR state, these are significantly better (within $6\%$ of the predicted value) than the results for quasihole exponents, likely because of the absence of (large) quasiparticles at the origin. Surprisingly, the Gf state also exhibits nice convergence of $g_{\text{h}}^{(1)}$ and $g_{\text{h}}^{(2)}$ toward each other and toward a value $g_{\text{h}} \approx 2.37$, which differs from the value $g_{\text{h}}=3$ predicted from CFT. Moreover, it violates ``spin-statistics,'' which requires bosons to have even integer values of $g$ (and non-unitary spin-statistics, which would allow odd integer values of $g$ for bosons).

We have provided a robust method of testing the screening properties of candidate FQH states, extracting the edge excitations' scaling exponents, and verifying the bulk-edge correspondence from microscopic model wavefunctions. Applying these methods for fundamental quasiholes and holes, we find that the Laughlin and MR states behave as they should for a properly screening FQH state, with scaling exponents matching those predicted for the expected edge CFT. The $\mathbb{Z}_3$-RR state exhibits stronger finite size effects for the quasihole exponent, though it appears to have proper screening and the hole exponent matches the CFT predicted value. In contrast, the quasihole and hole exponent computations provide a clear indication that the Gf wavefunctions do not exhibit the proper screening necessary for a legitimate FQH candidate. Despite this, the particle exponent computation indicates that there is some sort of screening occurring in the Gf state, which may explain why the ground state appeared to exhibit qualities similar to a FQH state, despite the fact that it, nonetheless, possesses pathologies.

\acknowledgements
We thank J. Dubail, D. Haldane, and especially R. Thomale for useful discussions. BAB and NR were supported by NSF CAREER DMR-095242, ONR - N00014-11-1-0635, Darpa - N66001-11-1-4110, Packard Foundation and Keck grant, and thank Microsoft Station Q for support and hospitality.


\end{document}